\documentstyle[aps,prd]{revtex}
\newcommand{\beq}{\begin{equation}}
\newcommand{\eeq}{\end{equation}}
\newcommand{\bea}{\begin{eqnarray}}
\newcommand{\eea}{\end{eqnarray}}

\newcommand{\ie}{{\em i.e.\/}}

\newcommand{\wh}{wormhole\/}

\begin{document}
\draft
\title{ On the stress-energy tensor of a rotating wormhole}
\author{S. E. Perez Bergliaffa\thanks{Electronic mail: \tt santiago@lafex.cbpf.br} and K. E. Hibberd}
\address{Centro Brasileiro de Pesquisas F\'{\i}sicas \\
Rua Dr.\ Xavier Sigaud 150, 22290-180 Urca, Rio de Janeiro, RJ -- Brazil} 
\date{\today}
\renewcommand{\thefootnote}{\fnsymbol{footnote}}
\twocolumn[
\hsize\textwidth\columnwidth\hsize\csname@twocolumnfalse\endcsname 
\maketitle

\begin{abstract}
\hfill{\small\bf Abstract}\hfill\smallskip
\par
We analyze the stress-energy tensor necessary to generate a general stationary and axisymmetric spacetime. The constraints on the geometry arising from 
considering a perfect fluid as a source are derived. For a fluid with a nonzero stress tensor, we obtain two necessary conditions on the metric.  
As an example, we show that the rotating wormhole presented in the literature can not be described by either a perfect fluid or by a fluid with anisotropic stresses.

\end{abstract}
\smallskip\mbox{}]
\footnotetext[1]{Electronic mail: \tt santiago@lafex.cbpf.br}
\renewcommand{\thefootnote}{\arabic{footnote}}

A  wormhole is a configuration of the gravitational and matter fields that can be understood as a ``handle'' connecting either two universes or two distant places in the same universe. Since the seminal paper by Morris and Thorne \cite{motho}, a number of static and dynamic wormholes have been found both in General Relativity and in alternative theories of gravitation 
(see \cite{hochvis1,hochvis2,nuestro} and references therein).  One of the most remarkable features of this field configuration is that under particular circumstances it permits the formation of closed timelike curves (CTC's) \cite{timemach}. 
Another striking characteristic of this type of spacetime is that 
it needs matter that violates the energy conditions as a source. 
This issue has been analyzed in detail both in static \wh s \cite{hochvis1} and in a completely general 
(\ie\/  non-symmetric and time dependent) traversable \wh\/ \cite{hochvis2,ida}.  It was shown that the null energy condition (NEC) must be violated in both cases. Let us remark that 
the issue of the existence of large amounts of NEC violating matter has not yet been settled.  
However, many results have been obtained assuming that wormholes exist and observational consequences of their existence have been examined
\cite{frono,frono2,gondiaz,cramer,torres,doqui2,kar1,kar2,ks1}.
These articles are concerned with the case of a static wormhole and many also assume that the configuration has spherical symmetry. Although these two hypotheses vastly simplify the calculations, it should be noted that astrophysical objects in nature are likely to have a nonzero angular momentum.  

Rotating wormholes were introduced by Teo \cite{teo}, who addressed the restrictions that must be imposed on the geometry in order to have a stationary and axisymmetric wormhole. 
However, in this article no consideration was given to the r.h.s. of Einstein's equations.  In other words, it is still not known what kind of matter could be a source of this otherwise geometrical construct.  In this brief report, we seek to shed light on this issue. We present the constraints on the Einstein tensor that arise from the matter used as a source of Einstein's equations for a general axially symmetric and stationary spacetime.  In turn, these constraints restrict the dependence of the metric coefficients on the coordinates.  
Finally, we examine the effect of these conditions on a particular geometry 
given as an example of a rotating wormhole in \cite{teo}.

The general metric for a stationary and axisymmetric spacetime can be written as \cite{hartle}
\begin{eqnarray}
ds^2 &=&- N^2dt^2 + e^{2\mu}dr^2  \nonumber \\
   &&+r^2 K^2 [d\theta^2 + \sin ^2\theta \;(d\phi - \omega\;dt)^2] ,
\label{metric}
\end{eqnarray}
where the metric coefficients $N,e^{2\mu}, K$ and $\omega$ depend only on $r$ and $\theta$. The function $\omega$ is the angular velocity of ``cumulative dragging'' \cite{hartle} and the function $K$ determines the proper radial distance.
In order to avoid a singularity on the axis of rotation, the $\theta $-derivative of the functions $N$, $\mu$, and $K$ must vanish for $\theta = 0, \pi$.

It is convenient to introduce the function $b=b(r,\theta )$, defined by
\beq
e ^{-2\mu} = 1-\frac b r .
\eeq
It is easily seen that with this definition, the metric Eq. (\ref{metric}) reduces to the metric introduced in \cite{motho} for the case of null rotation and spherical symmetry.
It was shown in \cite{teo} that for the metric given by Eq. (\ref{metric})  to describe a wormhole, the following conditions must be satisfied ($r_{\rm th}$ is the radius of the throat):

\begin{itemize}

\item $b$ must satisfy the inequality $b\leq r\;\forall r$, with the equality valid only at the throat,

\item $\frac{\partial b}{\partial r}|_{\rm r_{th}} < 1$ (``flaring out'' condition),

\item $\frac{\partial b}{\partial\theta}|_{\rm r_{th}} = 0$ (to ensure that the curvature scalar is nonsingular at the throat),

\item $N$ must be finite and nonzero for every $r$ (to ensure that no singularities or event horizons are present).

\end{itemize}
These are necessary (but not sufficient as we shall see below) conditions for the metric functions $N,b,K$ and $\omega$ to describe a rotating wormhole whose throat joins two identical regions. 

Notice that by defining the orthonormal tetrads

\begin{eqnarray}
\Theta^0 &= & N dt , ~~~ \Theta^1 = \left(1-\frac b r\right)^{-1/2} dr ,\nonumber \\
\Theta^2 &= & rK d\theta , ~~~ \Theta^3 = rK\sin\theta (d\phi - \omega dt) ,
\label{tetrad}
\end{eqnarray}
the metric Eq. (\ref{metric}) can be rewritten as
\beq
ds^2 = -(\Theta^0)^2 + (\Theta^1)^2 + (\Theta^2)^2 + (\Theta^3)^2 .
\label{eta}
\eeq

The energy-momentum tensor of a fluid with anisotropic stresses can be expressed as
\beq
T_{\mu\nu} = (\rho +p) u_\mu u_\nu + p\; \eta_{\mu\nu} + \Pi_{\mu\nu} ,
\label{t}
\eeq
where $\eta_{\mu\nu}$ is the metric given by Eq. (\ref{eta}) and $u_\mu$ is the four-velocity of the fluid. 
$\Pi_{\mu\nu}$ is the stress tensor and satisfies the constraints
\beq
\Pi^{\mu}_{\;\mu} = 0 ,
\label{ee7}
\eeq
and
\beq
\Pi_{\mu\nu}u^\nu = 0 .
\label{ee8}
\eeq
The one-form associated to the velocity must have the form
\beq
u_\mu = u_0\Theta^0 + u_3\Theta^3 ,
\label{vel}
\eeq
in order to respect the axisymmetry and stationarity of the spacetime described by Eq. (\ref{metric}). Its components are related by 
\beq
u_0^2 - u_3^2 =1 .
\label{v2}
\eeq
Einstein's field equations in the tetrad basis Eq. (\ref{tetrad}) are 

\beq
G_{00} = (\rho + p) u_0^2 - p +\Pi _{00} ,
\label{ee1}
\eeq
\beq
G_{33} = (\rho + p) u_3^2 + p + \Pi_{33} ,
\label{ee4}
\eeq
\beq
G_{03} = (\rho + p) u_0 u_3 + \Pi_{03} ,
\label{ee5}
\eeq
\beq
G_{11} = p+ \Pi_{11} ,
\label{ee2}
\eeq
\beq
G_{22} = p + \Pi_{22} ,
\label{ee3}
\eeq
\beq
G_{12} = \Pi_{12} ,
\label{ee6}
\eeq
and the remaining components of the Einstein tensor must be identically zero. 

We begin by considering a perfect fluid described by Eq. (\ref{t}) with $\Pi_{\mu\nu} = 0$. The system of equations  (\ref{v2}) - (\ref{ee6}) can be rearranged to the following

\beq
G_{11} - G_{22} = 0 ,
\label{c1}
\eeq
\beq
G_{12} = 0 ,
\label{c2}
\eeq
\beq
G_{03}^2 = (G_{00} + G_{22})(G_{11} - G_{33}) ,
\label{c3}
\eeq
\beq
u_0^2 = \frac{G_{00} + G_{22}}{G_{00}+G_{33}} ,
\label{c4}
\eeq
\beq
u_3^2 = \frac{G_{33}-G_{22}}{G_{00}+G_{33}} ,
\label{c5}
\eeq
\beq
\rho = G_{00} +G_{33} -G_{22} ,
\eeq
\beq
p = G_{22} .
\eeq
The first three equations are constraints to be satisfied by the Einstein tensor, while the remaining relations define the components of the velocity, energy density and pressure of the fluid.

The geometry described by Eq. (\ref{metric}) can have a perfect fluid as a source only if Eqs. (\ref{c1})-(\ref{c3}) are satisfied and the square of the components of the velocity Eqs. (\ref{c4})-(\ref{c5}) are non-negative.
It is easy to show that the metric given by Eq. (\ref{metric}) will not satisfy these constraints 
unless further restrictions on the metric coefficients are imposed.

We now consider a fluid with anisotropic stresses, for which 
Eqs. (\ref{v2})-(\ref{ee6}) constitute a system of ten equations in ten unknowns. This system can be conveniently rewritten in terms of the variable $x$ defined by

\beq
x = \frac{u_0}{u_3},
\eeq
which, after some algebraic manipulation, is given by the expression
\beq
x = \frac{G_{00}+G_{33}}{2G_{03}}\pm \frac{\sqrt{(G_{00}+G_{33})^2 - 4 G_{03}^2}}{2G_{03}} .
\label{xdef}
\eeq
The functions that solve the system and characterize the fluid are now given by
\beq
u_3^2 = \frac{1}{x^2-1},
\eeq
\beq
u_0 = x u_3 ,
\eeq
\beq
3p = G_{11} + G_{22} + G_{03} x - G_{00} ,
\label{een1}
\eeq
\beq
\rho = G_{03} x - G_{33} ,
\label{een2}
\eeq
\beq
\Pi_{33} = \frac{2p-G_{11} -G_{22}}{x^2-1} x^2 ,
\eeq
\beq
\Pi_{00} = G_{11} +G_{22} -2p + \Pi_{33} ,
\eeq
\beq
\Pi_{03} = \frac{\Pi_{33}}{x} ,
\eeq
\beq
\Pi_{11} =G_{11} - p ,
\eeq
\beq
 \Pi_{22} =G_{22} - p,
\eeq
\beq
\Pi_{12}=G_{12}   .
\eeq

The following restrictions must be imposed on the solution of the system:
\beq
G_{00}+G_{33}\geq 2 G_{03} ,
\label{r1}
\eeq
\beq
x^2 > 1.
\label{r2} 
\eeq
The first  constraint arises from Eq. (\ref{xdef}), while the second 
comes from Eq. (\ref{v2}) and restricts the Einstein tensor through Eq. (\ref{xdef}).
Let us emphasize that the metric given by Eq. (\ref{metric}) can be generated by a fluid with a nonzero $\Pi_{\mu\nu}$ only if conditions (\ref{r1}) and (\ref{r2}) are satisfied.

We now present an example to illustrate how
our constraints (16)-(18) and (35)-(36) limit the possible candidates for
the matter that generates a given metric.
Let us consider what type of matter corresponds to the particular example of a rotating wormhole given in \cite{teo}. The relevant functions are 
\beq
\omega = \frac{2a}{r^3}, \;\;\;
N = K = 1+\frac{(4a\cos\theta)^2}{r},\;\;\;
b = 1,
\label{metteo}
\eeq
where $a$ is the angular momentum of the wormhole. 

It was shown in \cite{teo} that these metric functions satisfy the geometric conditions needed in order to have a wormhole.  This geometry cannot be generated by a perfect fluid because Eq. (\ref{c2}) in this case gives
\begin{eqnarray}
16a^2 \cos^2\theta + 3 r = 0, \nonumber
\end{eqnarray}
which is not valid for every spacetime point. A fluid with anisotropic stresses cannot be a source of this wormhole either. This can be seen from Eq. (\ref{r1}) which may be rewritten in the form
\begin{eqnarray}
\sum_{n=0}^{6} q_n (\theta ) \;r^n \geq 0. \nonumber
\end{eqnarray}
This inequality will not hold for every value of $\theta$. Consequently
the source of the wormhole described by Eq. (\ref{metteo}) must 
incorporate more realistic (in the astrophysical sense) features, for example heat flux.

\vspace{.5cm}

{\bf Summary:}  We have derived the restrictions that a general axisymmetric and stationary spacetime described by the metric Eq. (\ref{metric}) must satisfy to have a perfect nonconvective fluid with differential rotation as a source. These restrictions are given by Eqns. (\ref{c1})-(\ref{c3}).  
An example of a metric that satisfies these constraints was given by 
Mars and Senovilla
\cite{mars1,mars2}, who found a family of axisymmetric and stationary spacetimes with a perfect fluid admitting a proper conformal motion as a source. It would be constructive to study if any of these solutions can represent a wormhole.

For a fluid with a nonzero stress tensor as
a source of Einstein's equations, we showed that the Einstein tensor, hence the metric given by Eq. (\ref{metric}) must satisfy Eqns. (\ref{r1}) and (\ref{r2}).

Let us remark that our results do not depend on the coordinate system chosen to describe the spacetime.  The fact that the tensor $G_{\mu\nu}-T_{\mu\nu}$ is zero for the metric Eq. (\ref{metric}) is a coordinate-independent statement. 

It must be noted that our results are valid for any stationary and axisymmetric spacetime. In particular, they can be applied to the spacetimes discussed in \cite{teo}. 
In that article, the conditions that must be imposed on the geometry in order to have a rotating wormhole were given.
However, it must be emphasized that an analysis of the matter side of Einstein's equations is essential. 
From our results it follows that 
the metric of a wormhole with a given type of fluid as a source must satisfy the corresponding constraints discussed above.
We demonstrated that the geometry proposed in \cite{teo} as a rotating wormhole can not be generated either by a perfect fluid or by a fluid with anisotropic stresses.

\vspace{0.5cm}
{\bf Acknowledgements:}  
SEPB acknowledges financial support from CONICET (Argentina). KEH is supported by FAPERj (Brazil).  The authors would like to thanks J. Salim for valuable discussions.  Many of the calculations in this paper were done with the package {\em Riemann} \cite{riemann}.

\end{document}